\documentclass[prd,preprint,eqsecnum,nofootinbib,amsmath,amssymb,preprintnumbers,longbibliography]{revtex4-1}

\usepackage{amsmath,graphicx}
\usepackage{color}
\usepackage[dvipsnames]{xcolor}
\usepackage[colorlinks,linkcolor=blue,citecolor=blue]{hyperref}
\usepackage[cal=boondox]{mathalfa}
\usepackage{slashed}
\definecolor{mygray}{gray}{0.6}

\def\eq{{\rm eq}}
\def\tr{{\rm tr}}
\def\p{{\boldsymbol p}}

\def\Eq#1{Eq.~(\ref{#1})}

\def\Fig#1{Fig.~\ref{#1}}

\def\Reff#1{Ref.~\cite{#1}}

\newcommand{\barr}[1]{\overline{#1}}
\newcommand{\f}{\mathcal{F}}

\def\bra{\langle}
\def\ket{\rangle}

\newcommand \beq{\begin{eqnarray}}
\newcommand \eeq{\end{eqnarray}}
\newcommand \be{\begin{equation}}
\newcommand \ee{\end{equation}}

\begin{document}

\title{Weak magnetic effect in quark-gluon plasma and local spin polarization }

\author{Jing-An Sun}
%\email{19110200014@fudan.edu.cn}
\affiliation{Institute of Modern Physics, Fudan University, Shanghai 200433, China}

\author{Li Yan}\email{cliyan@fudan.edu.cn}
\affiliation{Institute of Modern Physics, Fudan University, Shanghai 200433, China}
\affiliation{Key Laboratory of Nuclear Physics and Ion-beam Application (MOE), Fudan University, Shanghai 200433, China}

\date{\today}

\begin{abstract}

We propose the weak magnetic effect, which emerges in quark-gluon plasma close to local thermal equilibrium as the dissipative correction to the quark phase space distribution function, as a novel contribution to the observed Lambda hyperon local spin polarization. With a finite field strength, which is consistent with previous estimate of the magnetic field in heavy-ion collisions, one is able to explain the experimentally observed Lambda local spin polarization %of the second and the third order sine modulations, 
through all centrality classes.
Moreover, %our hydrodynamical simulaitons indicate that 
the weak magnetic effect plays an unambiguous role in the ordering between the second-order and the third-order modulations of the Lambda local spin polarization in experiments.

\end{abstract}
\maketitle

\section{introduction}

The thermodynamic characteristics %of an equilibrated quantum thermal system 
can be effectively inferred through experimental measurements once a system reaches local thermal equilibrium.
As can be seen from the equilibrium density operator, $\lg \hat \rho\sim \alpha \hat N - \beta_\mu \hat p^\mu + \omega_{\mu\nu} \hat M^{\mu\nu}/2$ (cf. Ref.~\cite{Becattini:2020sww}), the constants of motion $(\hat N, \hat p^\mu, \hat M^{\mu\nu})$, which correspond usually to the generators of a gauge symmetry group and the Poincar\'e group, are conjugate to the thermodynamic variables: chemical potential over temperature ($\alpha=\mu/T$), fluid four-velocity over temperature ($\beta_\mu = u_\mu/T$), and the thermal vorticity tensor ($\omega_{\mu\nu}=\omega^{\rm th}_{\mu\nu}\equiv-\frac{1}{2}(\partial_\mu \beta_\nu - \partial_\nu \beta_\mu)$). Given that these constants of motion are experimental observables, %independent of microscopic dynamics, 
through the measurements of %the conserved quantities, including 
charge, momentum and angular momentum of particles in the thermal system, one is allowed to extract
the information of chemincal potential, temperature, and so on.  

Such measurements have been carried out extensively, in particular in high-energy heavy-ion experiments. For instance, the observed hadron spectra -- momentum distribution -- 
%allows one to estimate 
reveals the information of temperature, baryon chemical potential and flow velocity in the quark-gluon plasma (QGP)~\cite{Andronic:2017pug}. Some more differential measurements, including the hadron collective flow and flow correlations, indicate that the space-time distribution of temperature in the QGP fireball is originated hydrodynamically from an almond-shape geometry~\cite{Ollitrault:1992bk,Shen:2020mgh,Heinz:2013th}. 

Recently, the study has been advanced to the analysis of hadron spin polarization, %and meson spin alignment, 
with the purpose of getting access to the rotational nature of QGP~\cite{Liang:2004ph,Betz:2007kg,STAR:nature,STAR:2018global,ALICE:2019global}. Indeed, with the assumption that QGP achieves local thermal equilibrium, the spin-1/2 particles are expected polarized owing to thermal vorticity~\cite{Becattini:2013fla,Becattini:2016gvu,Becattini:2020ngo,Huang:2011ru,Gao:2007bc}. Following the Cooper-Frye formula~\cite{Becattini:2013fla,Fang:2016CP},
\be
\label{eq:CF0}
P^\mu(\p) = - \frac{1}{8m} \epsilon^{\mu\alpha\beta\sigma} p_\sigma 
\frac{\int d\Sigma\cdot p n_F(1-n_F)\omega_{\alpha\beta}}{\int d\Sigma\cdot p n_F}\,,
\ee
with $n_F$ the Fermi-Dirac distribution, $\epsilon^{\mu\alpha\beta\sigma}$ the totally antisymmetric Levi-Civita symbol, and $\omega_{\mu\nu}=\omega^{\rm th}_{\mu\nu}$, the global spin polarization of Lambda hyperons emitted from a hyper-surface $\Sigma$ can be successfully characterized~\cite{STAR:nature,STAR:2018global,ALICE:2019global,Pang:2016global,Karpenko:2016global,Becattini:2016global,Ryu:2021chunshen}.

However, thermal vorticity fails to describe the differential spectrum of the Lambda spin polarization, which is often referred to as the local spin polarization (see \Reff{Becattini:2020ngo} for a recent review).  Especially, projected along the beam axis (z-axis), the thermal voticity yields an azimuthal distribution of the Lambda spin polarization that has an opposite sign comparing to experimental observations~\cite{Wu:2019localfail,Fu:2020oxj,Xia:2018tes,Becattini:2017global,STAR:2019local,ALICE:2021plocal}. The discrepency has intrigured extensive explorations of the spin generation from QGP beyond thermal vorticity and the local thermal equilibrium condition~\cite{PhysRevC.98.044906,PhysRevC.105.064901,Becattini:2021thermalshearcal,Hidaka:2017auj,LY:SIP,BBP:SIP, Wu:2022xiangyu,Yi:2021yichon,LY:SHE,Fu:2022SHE}.

In this Letter, %to resolve the sign problem of the local spin polarization, 
with respect to the conversion of quark spin from QGP fluid,
we introduce the dissipative effect induced by a weak external magnetic field. It has been noticed that, for the field strength $|eB|\ll m_\pi^2$, with $m_\pi$ the pion mass, while the bulk evolution of QGP is barely affected, quarks can be driven slightly out of local thermal equilibrium~\cite{Sun:2023pil,Sun:2023rhh}. The out-of-equilibrium effect leads to correction in the quark distribution function, which for quark species with electric charge $Q$, is~\cite{Sun:2023rhh}
\be
\label{eq:df0}
\delta f_{\rm EM}(x,\p) = n_F(1-n_F) \frac{\sigma_{\rm el}}{T \chi_{\rm ch}}
Q p_\mu F^{\mu\nu} \beta_\nu %\frac{q p_\mu F^{\mu\nu}  u_\nu}{(p\cdot u)^\delta}
\,,
\ee
where $\sigma_{\rm el}$ is the electrical conductivity, $\chi_{\rm ch}$ is an effective charge susceptibility, and %$E^\mu\equiv F^{\mu\nu}u_\nu$ defines the electric field in the fluid local rest frame. 
$F^{\mu\nu}$ the electromagnetic field strength tensor.
\Eq{eq:df0} is a scalar function solution with quark spin averaged. %degrees of freedom averaged. 
Correspondingly, when the weak magnetic field is taken into account, conversion of quark spin is modified, and the Cooper-Frye formula in \Eq{eq:CF0} would become
\be
\label{eq:df1}
P^\mu(\p) = - \frac{1}{8m} \epsilon^{\mu\alpha\beta\sigma} p_\sigma 
\frac{\int d\Sigma\cdot p \left[n_F(1-n_F) + (1-2n_F)\delta f_{\rm EM}\right]\omega_{\alpha\beta}}{\int d\Sigma\cdot p \left(n_F + \delta f_{\rm EM}\right)}\,.
\ee
In realistic heavy-ion collisions, since the magnetic field is created and orientated out of the reaction plane, the $p_\mu F^{\mu\nu}u_\nu$ factor in \Eq{eq:df0} results in a dipole structure in the azimuthal angle distribution. Therefore, the extra $\delta f_{\rm EM}$ does not contribute to the globle spin polarization,
%Under the weak field condition, $|\delta f_{\rm EM}|\ll n_F$, the induced correction gives rise to a small contribution to the quark spectrum. 
%However, 
in terms of local spin polarization, however, it is potentially significant. %could be sufficiently significant to flip the observed sign. 
\Eq{eq:df1} is the major result of this Letter.

Throughout the Letter, we use the natural units. 
Our convention of matrix is $g^{\mu\nu}=(+,-,-,-)$, and for the Levi-Civita symbol we take $\epsilon^{0123}=1$.

\section{Derivation of \Eq{eq:df1}}

In the presence of an %weak 
electromagnetic field, in QGP medium the evolution of quarks with spin degrees of freedom satisfies the Boltzmann-Vlasov equation~\cite{DeGroot:1980dk,PhysRevE.102.043203,PhysRevLett.59.1084},
\be
\label{eq:BV}
p^\mu \partial_\mu \f + Q F^{\mu\nu} p_\mu \frac{\partial \f}{\partial p^\nu} = - \mathcal{C}[\f] = -(p\cdot u) \frac{\f - \f_\eq}{\tau_R}\,.
\ee
Note that the quark distribution function is a Hermian matrix in the spin subspace, $\f=\f^\dagger$. 
%and spin average corresponds to $\frac{1}{2}\tr_2 \f$. 
%To facilitate analytical solutions, for 
For QGP close to local thermal equilibrium, we are allowed to linearize the collision kernal via the relaxation time approximation. In principle, relaxation time $\tau_R$ is a function of $p\cdot u$, depending on microscopic dynamics. In this Letter, we take the so-called quadratic ansatz in \Reff{Dusling:2009df}: $\tau_R = \bar{\tau} p\cdot u/T$, with $\bar\tau$ a free parameter to be determined by the transport coefficients. 
Because quarks in QGP are dominated by strong and electromagnetic forces, the collision kernel, as well as the parameter $\bar\tau$, are expected even with respect to the charge conjugation symmetry. Accordingly, subject to a charge conjugation transformation, \Eq{eq:BV} applies to anti-quarks. In the following, we shall focus on the derivation of quarks, while results associated with anti-quarks can be inferred upon charge conjugate symmetry.

By meanings of the Chapman-Enskog method, with the knowledge of the equilibrium quark distribution function~\cite{Becattini:2013fla}, 
 \be
 \label{eq:feq}
\f_{\rm eq} = \frac{1}{2m} \barr{U}(p) X(x,p) U(p)\,,
%\quad\barr{\f}_\eq = -\frac{1}{2m} (\barr{V}(p) \barr{X}(x,p) V(p))^T\,,
\ee
\Eq{eq:BV} can be solved order by order~\cite{DeGroot:1980dk,Sun:2023rhh}.
In \Eq{eq:feq}, $U(p)$ is the amalgamated spinor solution for free fermions with both spin up and spin down, $U(p) = (u_+(p), u_-(p))$, and $\barr{U}(p)= U^\dagger \gamma^0$. The matrix $X(x,p)$ is identified as
$
X(x,p) \equiv \left(e^{\beta\cdot p} \exp{\left[-\frac{1}{2}\omega_{\mu\nu}\Sigma^{\mu\nu}\right]}+1\right)^{-1}\,,
%\quad
%\barr{X}(x,p) = \frac{1}{e^{\beta_\mu p^\mu} \exp{\left[\frac{1}{2}\omega_{\mu\nu}\Sigma^{\mu\nu}\right]+1}}\,,
$
with $\Sigma^{\mu\nu}=\frac{i}{4}[\gamma^\mu,\gamma^\nu]$ the generator of relativistic rotation realized through the gamma matrix.
At the leading order of $|eB|/T^2$, the solution can be found as, $\f = \f_\eq + \delta \f_{\rm EM}$, where the dissipative correction is
\be
\label{eq:dfem1}
\delta \f_{\rm EM}
= - \frac{\bar\tau}{T} Q F^{\mu\nu} p_\mu \frac{\partial}{\partial p^\nu} \f_\eq
%= -\frac{1}{2m} \frac{\tau_R}{p\cdot u} q F^{\mu\nu} p_\mu \frac{\partial}{\partial p^\nu}\left(\bar U(p) X U(p)\right)
= \frac{1}{2m} \barr{U}(p) Y(x,p) U(x,p)\,,
%= -\frac{\tau_R}{2m} %\frac{\tau_R}{p\cdot u} 
% Q F^{\mu\nu} p_\mu 
%\left(\bar U(p)  \frac{\partial X}{\partial p^\nu} U(p)\right)
\ee
with 
\be
Y(x,p)\equiv - \frac{\bar\tau}{T} Q F^{\mu\nu} p_\mu \frac{\partial X}{\partial p^\nu}
=   \frac{\bar\tau}{T} Q F^{\mu\nu} p_\mu\beta_\nu e^{\beta\cdot p}X^2  \exp{\left(-\frac{1}{2}\omega_{\alpha\beta}\Sigma^{\alpha\beta}\right)} \,.
\ee
In deriving \Eq{eq:dfem1}, we have used the Dirac equation $(\slashed{p}-m)U(p)=0$, which implies that $F^{\mu\nu}p_\mu \frac{\partial}{\partial p^\nu} U(p)=0$ and $F^{\mu\nu}p_\mu \frac{\partial}{\partial p^\nu} \barr{U}(p)=0$. Hermiticity of $\delta \f_\eq$ can be verified, with respect to the condition $Y^\dagger = \gamma^0 Y \gamma^0$. 

In accordance with the dissipative correction in the conserved current of hydrodynamics, the dissipative correction in the phase-space distribution function satisfies the Landau's matching condition. Especially, for a QGP medium with local charge neutrality condition, the leading order dissipative correction to the current is driven by the external electromagnetic field, as,
\be
\label{eq:current}
\Delta j^\mu = \sigma_{\rm el} E^\mu = \sum Q_f \int \frac{d^3 \p}{p^0} p^\mu \tr_2[\delta \f_{\rm EM} - \delta \barr{\f}_{\rm EM}]\,, 
\ee
where the trace, $\tr_2$, is taken in the spin subspace and the summation takes into account of flavors and colors.  %with respect to spin indices. 
To evaluate the trace, using the cyclicity in a trace and $U(p)\barr{U}(p)=\sum_{\rm spin} u(p) \bar u(p)=\slashed{p}+m$, and the property that trace of three gamma matrices vanishes, one finds for quarks with electric charge $Q$,
\begin{align}
\label{eq:tr2}
\tr_2 \delta \f_{\rm EM} &= \frac{1}{2} \tr_2 Y = \frac{\bar\tau}{2T} Q F^{\mu\nu} p_\mu\beta_\nu e^{\beta\cdot p} \tr_2\left[ X^2  \exp{\left(-\frac{1}{2}\omega_{\alpha\beta}\Sigma^{\alpha\beta}\right)} \right] \cr
& = \frac{\bar\tau}{2T} Q F^{\mu\nu} p_\mu\beta_\nu e^{\beta\cdot p} 
\sum_{m,n=1} (-1)^{m+n+2} e^{-(m+n)\beta\cdot p}\tr_2\left[\exp\left(\frac{m+n-1}{2}\omega_{\alpha\beta}\Sigma^{\alpha\beta}\right)\right]\cr
& =  -\frac{\bar\tau}{2T} Q F^{\mu\nu} p_\mu\beta_\nu (4 n_F' + \frac{ \omega_{\alpha\beta}\omega^{\alpha\beta}}{2} n_F''')\,.
%\cr
%& = 2 \delta f_{\rm EM} -\frac{\bar \tau_R}{4T} QF^{\mu\nu} p_\mu \beta_\nu n_F'''  \omega_{\alpha\beta}\omega^{\alpha\beta}
\end{align}
In the above expression, prime over $n_F$ indicates derivative with respect to $p\cdot u/T$, hence for instance, $n_F'=-n_F(1-n_F)$. A couple of comments are in order. First, we have used the approximated identity in the derivation~\cite{Becattini:2013fla}: $\tr_2\left[\exp\left(\frac{n}{2}\omega_{\alpha\beta}\Sigma^{\alpha\beta}\right)\right] \approx 4 + \frac{n^2}{2} \omega^{\alpha\beta}\omega_{\alpha\beta}$. Secondly, substituting the trace back to the matching condition \Eq{eq:current} yields an expression that relates $\bar\tau$ to the electrical conductivity $\sigma_{\rm el}$. In the limit $\omega\to 0$, one arrives at the standard relation $\sigma_{\rm el} = \bar\tau\chi_{\rm el}$~\cite{Sun:2023rhh}. Accordingly, up to a factor of two that amounts to a summation over spin, one identifies the first term in \Eq{eq:tr2} as the scalar function $\delta f_{\rm EM}$. In sum, as expected, in a non-rotating system the spin averaged quark distribution function reduces to $\frac{1}{2}\tr_2 \f=n_F + \delta f_{\rm EM}$.

The Cooper-Frye formula in \Eq{eq:CF0} is a variant of the the Pauli-Lubanski pseudo-vector, 
\be
\label{eq:PL}
\Pi_\mu^{\rm PL} = -\frac{1}{2m}\epsilon_{\mu\rho\sigma\tau} p^\tau S^{\rho\sigma} \,,
\ee
where $S^{\rho\sigma}$ corresponds to the angular momentum operator. Spin polarization of particles emitted from QGP, in particular, can be described %via the Pauli-Lubanski pseudovector 
when $\Pi^{\rm PL}_\mu$ is evaluated on a hypersurfce that meats the freeze out condition in heavy-ion collisions, and as the averaged expectation of particles in phase space.
%which represents the averaged spin of particles being emitted from a hypersurface. %With the weak magnetic correction $\delta \f_{\rm EM}$, it receives extra contribution. From 
As has been shown previously~\cite{Becattini:2013fla}, %for spin polarization 
because $S^{\mu\nu}$ receives contributions only from spin, which with respect to the quark distribution function in QGP out of local thermal equilibrium, can be solved from the spin density tensor~\cite{DeGroot:1980dk},
\begin{align}
s^{\lambda,\rho\sigma}(x) &= \frac{1}{2}\int \frac{d^3 \p}{2p^0} \tr_2\left(\f\barr{U}(p)\{\gamma^\lambda,\Sigma^{\rho\sigma}\}U(p)\right) %- \mbox{anti-quark}
 = \int\frac{d^3 \p}{2p^0}\left[p^\lambda \Theta^{\rho\sigma} + p^\rho \Theta^{\sigma\lambda} + p^\sigma \Theta^{\lambda\rho}\right]\,,
\end{align}
where the rank-2 tensor function is,
\be
\label{eq:Theta}
\Theta^{\mu\nu}(x) \equiv \tr_2\left[(X+Y)\Sigma^{\mu\nu}\right]
%=n_F'\omega^{\mu\nu} + n_F'' \frac{\sigma_{\rm el}}{T\chi_{\rm el}} QF^{\alpha\beta}p_\alpha \beta_\beta \omega^{\mu\nu} 
= \left[n_F(1-n_F) + (1-2n_F) \delta f_{\rm EM}\right]\omega^{\mu\nu}\,.
\ee
Accordingly, one finds average with respect to quark spectrum,
\be
\label{eq:Smn}
P_\mu=
\bra \Pi_\mu^{\rm  PL}\ket 
=-\frac{1}{2m} \epsilon_{\mu\rho\sigma\tau} p^\tau
\frac{1}{\tr_2 \f} \frac{d s^{0,\rho\sigma}}{d^3 \p} = 
-\frac{1}{4m} \epsilon_{\mu\rho\sigma\tau} p^\tau \frac{\Theta^{\rho\sigma}}{\tr_2 \f}\,.
\ee 
Following the Cooper-Frye prescription~\cite{Cooper:1974mv} and integrate over a hypersurface, Eqs.~(\ref{eq:Theta}) and (\ref{eq:Smn}) lead to \Eq{eq:df1}. 
Note that the derivation does not reply on the the explicit form of $\omega^{\mu\nu}$, hence
\Eq{eq:df1} applies as well when other contributions are taken into account beyond the thermal vorticity tensor, including the shear-induced polarization.

\section{Local spin polarization in heavy-ion collisions}

In realistic heavy-ion collisions, in the collision region there must be an electromagentic field created pointing out of reaction plane, % as the two nuclei passing through each other, 
namely, ${\bf B} = B_y\hat y$.  Although the field strength is expected strong initially, %at the instant of the collision, 
it drops drastically~\cite{Deng:2012pc,Bzdak:2011yy,Skokov:2009qp,Tuchin:2013apa,McLerran:2013hla,Hattori:2016emy,Yan:2021zjc,Huang:2022qdn}. It is very likely that the magnetic field becomes weak when the QGP medium is formed and starts to expand hydrodynamically~\cite{Yan:2021zjc,Huang:2022qdn}. 
%Let us emphasize that the weak magnetic field we are considering is in the context of QCD scales, in particular, $|eB|\ll m_\pi^2$, yet it could still be an ultrastrong field in nature. 
Although the magnetic effect in QGP is too weak to influence the bulk medium evolution, it has been found significant to some of the particular signatures, including the anisotropy in the thermal photon spectrum~\cite{Sun:2023pil,Sun:2023rhh} and the thermal di-lepton polarization~\cite{Wei:2024}.

In the presence of a weak magnetic field,
%When spin degrees of freedom is taken into account, 
as indicated by \Eq{eq:df1}, although the physics that quarks get polarized due to the rotation in QGP is not modified, %in the presence of a weak magnetic field 
the quark spin %polarization 
configuration in phase space can be dramatically different. To show this, we apply \Eq{eq:df1} to the hydrodynamical modeling of QGP evolution, and calculate the spin polarization of s-quarks emitted from the chemincal freeze-out hypersurface. Because the spin of a Lambda hyperon is dominantly carried by the s-quark, as u-, d- and s-quark coalesce~\cite{Sheng:2020ghv}, the calculation is also expected a good approximation to the polarization of Lambda hyperon. Of course, as a consequence of the approximate chiral symmetry breaking during the coalescence, %instead of taking the current quark mass or valance quark mass, %the mass of the s-quark in the Lambda hyperon differs from its current quark mass. We shall 
we leave the s-quark mass $m_s$ as a free parameter. % for the calculation of the Lambda polarization.

With respect to the isobar collision systems with $\sqrt{s_{NN}}=200$ GeV, we carry out hydrodynamical simulations %for the system expansion 
using the 3+1D MUSIC program~\cite{Schenke:2010rr}. Initial conditions are chosen and calibrated, according to the centrality class determined by the charged particle multiplicity and a longitudinal profile that %has quantitatively reproduced
reproduces the rapidity dependent direct flow~\cite{Bozek:2015bna}. In order to capture the thermal vorticity associated with the elliptic and triangular initial geometry, we deform the optical Glauber model distribution with a finite ellipticity and triangularity, following the prescription introduced in \Reff{Teaney:2010vd}. 
%More details on our initial condition can be found in \Reff{}.
In addition to the standard set of parameters that has been used in the hydrodynamical modeling, % and successfully described various collective flow observables, 
%The equation of state extracted from Lattice QCD, and transport coefficients 
such as the shear and the bulk viscosities, 
we choose the mean value of electrical conductivity $\sigma_{\rm el}/T = 1$, with respect to the pQCD expectation $\sigma_{\rm el}/T\in [0.2,2]$~\cite{Arnold:2000dr,Floerchinger:2021xhb}. To avoid the complexity from the unknown space-time evolution, for the magnetic field we take ${\bf B}= B\Gamma(\eta_s)\hat y$, with $B$ a constant parameter and the space-time rapidity dependence $\Gamma(\eta_s)$ determined from the retarded potential solution~\cite{Hattori:2016emy}. In this way, $|eB|$ is the time-average field strength in the center of the QGP.
%are implemented in the current study. 

\begin{figure}
\begin{center}
\includegraphics[width=0.75\textwidth] {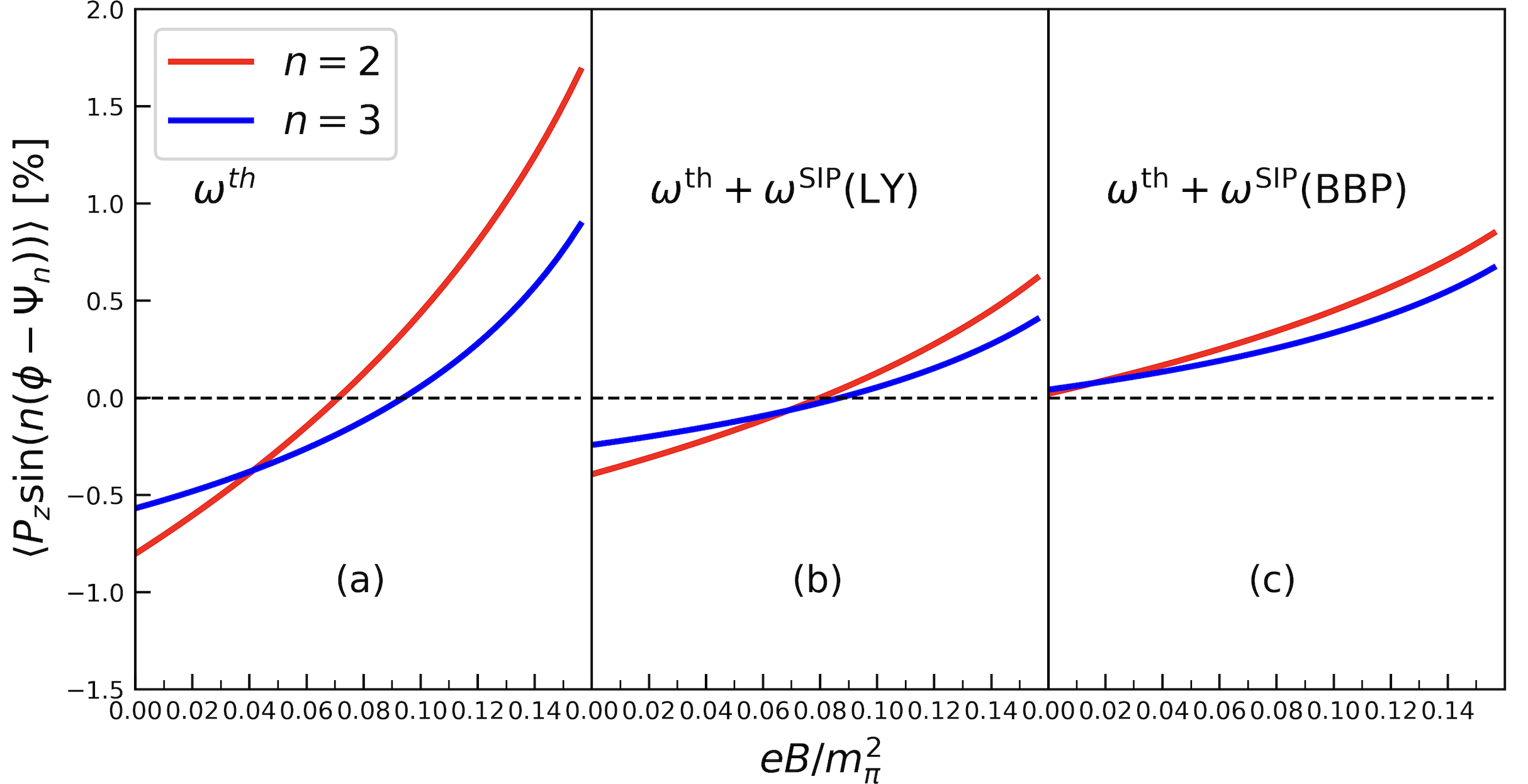}
\caption{
\label{fig:fig1} Dependence of the second-order modulation (red) and the third-order modulation (blue) of Lambda local spin polarization on the magnetic field.
}
\end{center}
\end{figure}

Lambda local spin polarization along the beam axis can be quantified when it is decomposed into sine Fourier modulations. Under the local  thermal equilibrium condition, the local thermal vorticities in QGP %the second order and the third order harmonics 
are expected from the medium response to the elliptic and the triangular geometries, respectively. Shown in \Fig{fig:fig1}, we solve the Lambda local spin polarization for RuRu collisions with impact parameter $b=7$ fm and $m_s=0.8$ GeV. Indeed, in \Fig{fig:fig1} (a) where $w_{\mu\nu}$ is identified as $w^{\rm th}_{\mu\nu}$, without magnetic field one finds negative second-order and third-order modulations of the local spin polarization. %$n=2$ and $3$. 
It is interesting to notice that the magnitude for the second order %$n=2$ 
is larger, in consistency with the hierarchy between elliptic flow and triangular flow. %the fact that elliptic flow is normally larger than triangular flow. 

In order to resolve the sign problem of the Lambda local spin polarization, a number of contributions have been proposed beyond thermal vorticity~\cite{Hidaka:2017auj,LY:SIP,BBP:SIP,Yi:2021yichon,Wu:2022xiangyu}. Of particular interest, the shear tensor $\xi_{\mu\nu} = \frac{1}{2}(\partial_\mu \beta_\nu + \partial_\nu \beta_\mu)$, which appears in systems in local thermal equilibrium, gives rise to the so-called shear-induced polarization $\omega^{\rm SIP}_{\mu\nu}$. Corresponding to the two types of formulation, %for the shear-induced polarization, 
$\omega^{\rm SIP}_{\mu\nu}({\rm LY})$ by Liu and Yin~\cite{LY:SIP}, and $\omega^{\rm SIP}_{\mu\nu}({\rm BBP})$ by Becattini, Bucciantini and Palermo~\cite{BBP:SIP}, with the replacement $\omega_{\mu\nu}\to \omega^{\rm th}_{\mu\nu} + \omega^{\rm SIP}_{\mu\nu}$ in \Eq{eq:df1}, we re-calculate the Lambda local spin polarization. Results are shown in \Fig{fig:fig1} (b) and (c).

Let us simply summarize our observations from \Fig{fig:fig1}. First, in all cases, with the magnetic field introduced, both modulations in the local spin polarization get increased monotonically. Eventually,  with a weak magnetic field, they become positive %$|eB|\lesssim 0.1 m_\pi^2$, 
regardless of the contribution from the shear-induced polarization. Secondly,
%With also the shear-induced polarization, %similarly, both modes in the local spin polarization increase monotonically with respect to the magnetic field, but 
the second-order modulation %$n=2$ mode 
appears more sensitive to the magnetic field than the third-order one, %$n=3$ mode, 
while the sensitivity is %apparently 
reduced by the shear-induced polarization.
Lastly, and quite importantly, although %our hydrodynamic simulations confirm the role of the shear-induced polarization that it 
the shear-induced polarization can make the local Lambda spin polarization positive%, as seen in \Fig{fig:fig1} (c)
~\cite{Alzhrani:2022dpi}, the ordering that the second-order modulation is greater than the third-order one, %$n=2$ mode is larger than $n=3$, 
is an unambiguous signature associated with the magnetic field.  
All of these qualitative features are not affected by the s-quark mass, though for smaller $m_s$, we do find that the spin conversion from the QGP is earsier, % to be realized, 
which effectively reduces the required field strength.

\begin{figure}
\begin{center}
\includegraphics[width=0.8\textwidth] {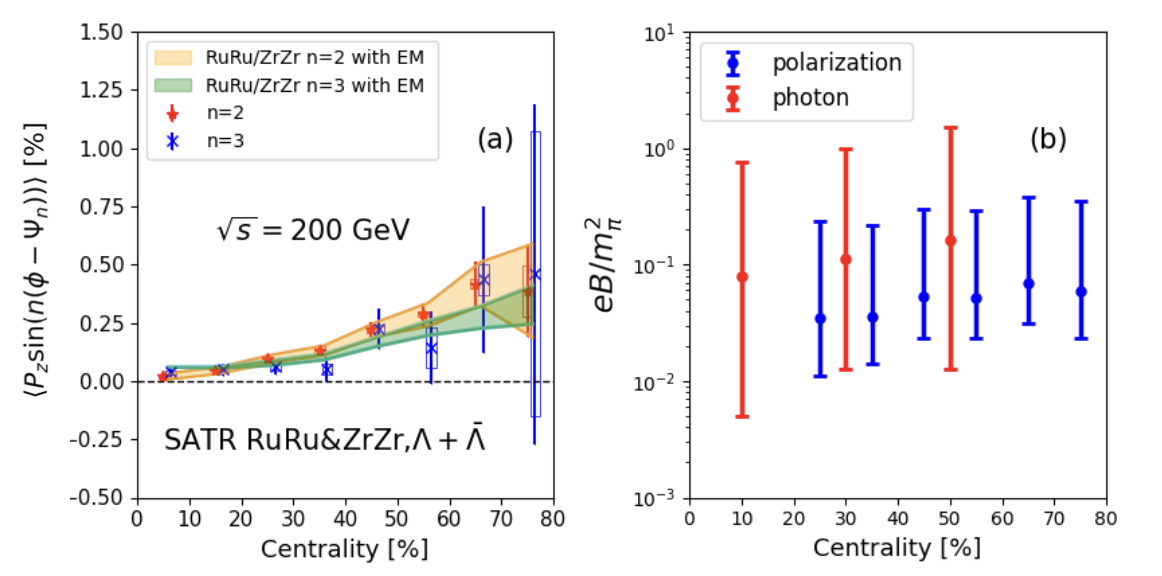}
\caption{
\label{fig:fig2} (a) Centrality dependent Lambda local spin polarization from the isobar collisions: symbols are experimental results from the STAR collaboration~\cite{STAR:2023local} while % dashed lines and 
shaded band correspond to theoretical results %without and 
with magnetic field for $\omega=\omega^{\rm th} + \omega^{\rm SIP}$(BBP), respectively. (b) Centrality dependence of the magnetic field extracted according to the Lambda local spin polarization (blue symbols) and the direct photon elliptic flow (red symbols)~\cite{Sun:2023rhh}.
}
\end{center}
\end{figure}

The experimental observables, especially the centrality dependent %of $n=2$ and $n=3$ modes of the 
Lambda local spin polarization can be well captured in the presence of the weak magnetic field. Although our current calculation is approximate, we achieve excellent characterizations %can be achieved 
with $m_s=0.8$ GeV and $w_{\mu\nu} = \omega^{\rm th}_{\mu\nu}+w^{\rm SIP}_{\mu\nu}$(BBP). 
% would not be a surprise that the experimentally measured Lambda local spin polarization can be described with respect to a specified field strength. %Especially, as centrality increases the field strength grow due to the fact there are more charged spectators participating the generate of the magnetic field, which explains naturally the systematic increase of the local spin polarization. 
Shown in \Fig{fig:fig2} (a), the STAR measured $n=2$ modulation of the Lambda local spin polarization in the isobar systems can be reproduced~\cite{STAR:2023local}, with the the centrality dependent $|eB|$ extracted and given in \Fig{fig:fig2} (b) (blue symbols). Error bars in \Fig{fig:fig2} (b) stem from the experimental uncertainties, as well as $\sigma_{\rm el}/T\in [0.2,2]$. For comparison, in \Fig{fig:fig2} (b), the estimated field strength from the direct photon $v_2$ in AuAu collisions are shown as well~\cite{Sun:2023rhh}. 
It is interesting to notice that the extracted field strength grows as one proceeds towards peripheral collisions.
With the same field strength, %the results of the $n=3$ mode 
the third-order modulation of the Lambda local spin polarization can be described as well, with the results shown as the green shaded band, in \Fig{fig:fig2} (a).

\begin{figure}
\begin{center}
\includegraphics[width=0.5\textwidth] {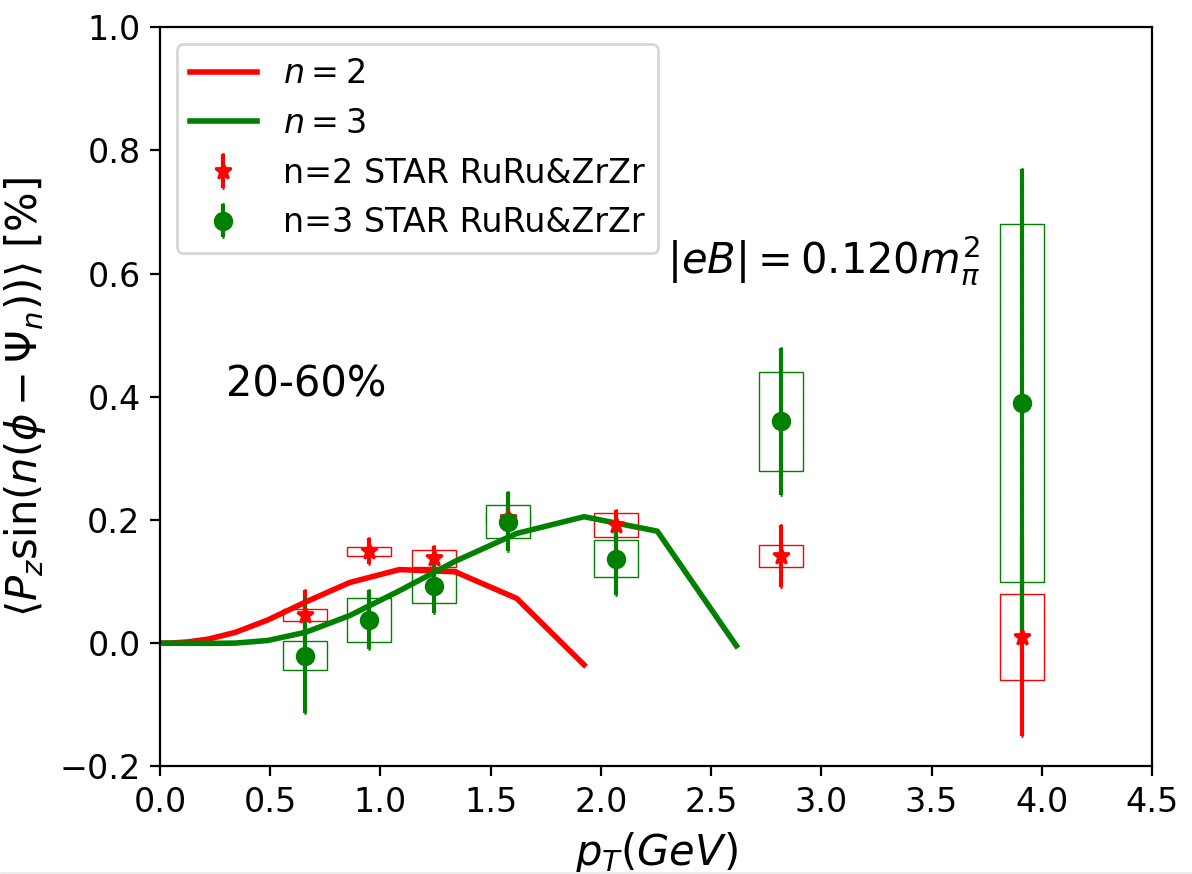}
\caption{ Transverse momentum dependent Lambda local spin polarization for the $n=2$ (red) and $n=3$ (green) modes of the isobar collisions in the centrality class 20\%-60\%. 
\label{fig:fig3} 
}
\end{center}
\end{figure}

In the centrality class 20\%-60\%, the STAR collaboration also measured the transverse momentum dependent local spin polarization for the Lambda hyperon~\cite{STAR:2023local}. Although the ordering in \Fig{fig:fig2} (a) is not obvious, as shown in \Fig{fig:fig3}, from the $p_T$ dependent %modes of the 
Lambda local spin polarization, in the low $p_T$ region one clearly observes a larger $n=2$ modulation than $n=3$. As we emphasized previously, this ordering indicates the presence of a weak magnetic field. Moreover, our theoretical calculation gives consistent description of the $p_T$ dependence. The deviations at large $p_T$, where hydrodynamics becomes invalid and perturbative contributions to spin local polarization can be important, should not be a surprise.

\section{Summary and discussion}

For the conversion of quark spin from QGP, and consequently the spin polarization of Lambda hyperon, we find a novel effect originated from the weak magnetic field. 
Unlike the coupling via a finite magnetic moment of the hyperon $\mu_\Lambda$, namely, with $\omega_{\mu\nu}\to \omega_{\mu\nu} + Q F_{\mu\nu} \mu_\Lambda/T$~\cite{Becattini:2016gvu}, this novel effect reflects a small dissipative correction 
%Although it yields a small correction 
in the quark phase space distribution. The weak magnetic effect can be crucial to the observed Lambda local spin polarization, not only because it correctly captures the sign %of the local polarization 
observed in experiments, but also it is responsible to %generates naturally 
the ordering between the second- and the third-order modulations of the local spin polarization. Using hydrodynamical simulations, the experimentally measured Lambda local spin polarization can be well understood. The required fields in different centralities, as shown in \Fig{fig:fig2} (b), are found consistent with the estimation made based on the direct photon elliptic flow. These are indeed weak fields, as they satisfy $|eB|\ll m_\pi^2$. In fact, using the mean value from \Fig{fig:fig2} (b), % as $|eB|\sim 0.1 m_\pi^2$, 
one finds the time integrated value ${\mathcal{B}}=\int dt eB\lesssim 50$ MeV, which is too weak to induce the splitting in the global polarization between $\Lambda$ and $\bar \Lambda$~\cite{Guo:2019joy}.

%The weak magnetic effect in the current work does not contribute to the global spin polarization, as $\delta f_{\rm EM}$ has naturally a dipole structure which vanishes after angle integration. 
%In principle, weak magnetic field can contribute alternatively as, $\omega_{\mu\nu}\to \omega_{\mu\nu} + Q F_{\mu\nu} \mu_\Lambda/T$, which leads to splittings in the global polarization between $\Lambda$ and $\bar \Lambda$.

\acknowledgements

We are grateful to Xiangyu Wu for providing us the centrality classifications with eccenticity parameters for the isobar systems. This work is supported by the National Natural Science Foundation of China under Grant No. 11975079 and 12375133.

%%%%%%%%%%%%%%%%%%%%%%%%%%%%%%%%
\bibliography{references}

\end{document}